\documentclass[a4paper, 12pt]{article}      
\usepackage{a4}
\usepackage{amsfonts}
\usepackage{amssymb}
\usepackage{epsfig}

\author{}

\newcommand{\be}{\begin{equation}}
\newcommand{\ee}{\end{equation}}
\newcommand{\ba}{\begin{array}}
\newcommand{\ea}{\end{array}}
\newcommand{\bea}{\begin{eqnarray}}
\newcommand{\eea}{\end{eqnarray}}

\def\IR{\relax{\rm I\kern-.18em R}}

\def\IP{\relax{\rm I\kern-.18em P}}
\def\inbar{\vrule height1.5ex width.4pt depth0pt}
\def\IC{\relax\,\hbox{$\inbar\kern-.3em{\rm C}$}}

\def\K3{{\bf K3}}

\def\n2d{\cN_{V^*}^{\otimes 2}}

\def\IC{\mathbb{C}}

\def\IR{\mathbb{R}}

\def\IP{\mathbb{P}}

\def\cN{{\mathcal N}}

\begin{document}

\title{
\begin{flushright} \vspace{-4cm}
{\small CERN-PH-TH/2009-243\\
\small MPP-2009-205 \\
 \small LMU-ASC 56/09\\
\vspace{-0.35cm}
hep-th/yymmnnn} \end{flushright}
\vspace{0.5cm}
 Evaporation of Microscopic Black Holes in String Theory and the Bound  on Species} 

\date{}

\maketitle


\vspace{-1.4cm}


\begin{center}

{\bf Gia Dvali}$^{a,b,d,}$\footnote{georgi.dvali@cern.ch} and {\bf Dieter L\"ust}$^{c,d,}$\footnote{dieter.luest@lmu.de, luest@mppmu.mpg.de}

\vspace{.6truecm}

{\em $^a$CERN,
Theory Department\\
1211 Geneva 23, Switzerland}

\vspace{.2truecm}

{\em $^b$Center for Cosmology and Particle Physics\\
Department of Physics, New York University\\
4 Washington Place, New York, NY 10003, USA}

\vspace{.2truecm}

{\em $^c$Arnold Sommerfeld Center for Theoretical Physics\\
Department f\"ur Physik, Ludwig-Maximilians-Universit\"at M\"unchen\\
Theresienstr.~37, 80333 M\"unchen, Germany}

\vspace{.2truecm}

{\em $^d$Max-Planck-Institut f\"ur Physik\\
F\"ohringer Ring 6, 80805 M\"unchen, Germany}

\end{center}

\vspace{0.5cm}

\begin{abstract}
\noindent  
 We address the question
how string compactifications with D-branes are consistent with the black hole bound, which arises in any theory with number of particle species to which the black holes can evaporate. 
  For the  Kaluza-Klein particles,  both longitudinal and transversal to the D-branes, it is
relatively easy to see that the black hole bound is saturated, and the geometric relations can be understood in the language of species-counting.   
 We next address the question of the black hole evaporation into the higher string states and discover, 
 that contrary to the naive intuition,  the exponentially growing number of Regge  states does not preclude the existence of semi-classical black holes of sub-stringy size.  Our analysis indicates that  the effective number of string resonances to which such micro black holes  evaporate  is not exponentially large but is bounded by $N = 1/g_s^2$, which suggests the interpretation of the well-known relation between the Planck and string scales as the saturation of the black hole bound  on the species number.   
In addition, we also discuss some other issues in D-brane compactifications with a low string scale of order TeV, such as the masses of light moduli fields.

\end{abstract}

\thispagestyle{empty}       
\clearpage

\tableofcontents

\section{Introduction}

String theory, being the theory of one-dimensional extended objects, contains an infinite
number of elementary particles, when seen from the field theory point of view.  Specifically, the excitation spectrum of
a relativistic string contains  a tower of infinitely many higher spin excitations, whose number grows exponentially
with the masses of the excited string modes. In fact, these states follow the well-known Regge excitation pattern with masses
\begin{equation}
M_n^2=n M_s^2\, ,
\end{equation}
where $M_s$ is the intrinsic string scale ($M_s^{-2}=\alpha'$), and $n$ counts the n$^{th.}$ excitation level.

In string compactifications from ten to four space-time dimensions there are several additional generic mass scales and  particles, which are
related to the internal geometry of the compact space.
These are Kaluza-Klein (KK) particles, whose masses scale inversely with some internal radii, possible winding states, as well as moduli
fields that determine the parameters of the underlying geometry. In particular, in intersecting D-brane compactifications, there are several distinct mass scales and moduli fields, which could be also of experimental
relevance. Basically, each of these mass scales is related to the Compton wave-lengths of some particles,
which generically arise during the compactification process.

A particularly interesting scenario is given by D-brane compactifications with a string scale $M_s$ being near the scale
of the Standard Model (SM), i.e. $M_s={\cal O}({\rm TeV})$ \cite{Antoniadis:1998ig}, which implies that some of the extra dimensions must be much larger than
$M_s^{-1}$. Moreover, since $M_s$ sets the fundamental scale of gravity in the higher dimensional space, the gravitational effects will start to
play an important role already at low energies, and possibly be accessible experimentally. In fact, in this scenario there is an explosion of string states at the TeV scale,
and some spectacular model-independent stringy signatures are expected at the LHC in case the string scale is low
\cite{Lust:2008qc,Anchordoqui:2008di,Anchordoqui:2009mm}. 
D-brane compactifications with a low string scale are indeed possible, because the open string gauge interactions are confined to the lower dimensional world volumes
of the D-branes.   On the other hand,  the gravitational force is enormously weakened by the fact, that the closed string gravitons
can leak out into the large transversal bulk space.
Of course, one has to investigate, if low strings scale D-brane compactifications
can be dynamically realized by a certain moduli potential, a question, which is not a focus of our paper, but which we shall only briefly touch later.

Low string scale D-brane models are a concrete, microscopic  realization of the large extra-dimension scenario of ADD,
which was originally introduced as a phenomenologically-motivated solution  to  the hierarchy
problem \cite{ArkaniHamed:1998rs}. Of course, in string theory compactifications, some new phenomenological questions 
compared to ADD arise, such as the appearance of several new mass scales, the existence of additional states and in particular
the problem of light moduli at tree level, which typically arise in case of a large internal string manifold. We will discuss the light moduli
problem and radiative mass corrections to the moduli fields later on.

  The main focus of the present paper is to investigate how the string theory
compactifications, viewed as the theories of many particle species, cope with the consistency requirements of the semi-classical black hole (BH) physics \cite{Dvali:2007hz, Dvali:2008sy, cesar, ramy}.  
  
  This question emerges already at the level of high-dimensional field theories. Indeed, from the point of view of a four-dimensional observer, any high-dimensional theory 
   is a theory of large number of particle species, since every high-dimensional field (e.g., a graviton)  produces a tower  of  KK excitations.  An each member of this tower represents  a four-dimensional particle with a definite mass and other quantum numbers. 
    The intrinsic geometric property of  large extra dimensional scenarios is the hierarchy between the 
 four-dimensional  ($M_4$) and high-dimensional ($M_{4+d}$) Planck scales produced by  the large compactification volume
 \begin{equation} 
   M_4^2 \, = \, M_{4+d} ^2   (M_{4+d}^dV^{(d)}) \, ,   
 \label{mplancks} 
 \end{equation}
 where the quantity $M_{4+d}^dV^{(d)}$ is the volume of the compact manifold measured in the units of 
 the high-dimensional Planck length $l_{4+d} \, \equiv \, M_{4+d}^{-1}$.    
 
  Recently \cite{Dvali:2007hz, Dvali:2008sy, cesar, ramy}, it was understood that the above simple geometric relation, is in fact a particular 
  realization of a much more general bound, according to which in any theory with $N$ particle species, there is  an inevitable hierarchy between the Planck mass and the fundamental gravitational cutoff 
  of the theory, which we shall denote by  $M_*$.  Among several possible equivalent definitions of 
  $M_*$, we shall use the one  in which $l_*  \, \equiv\, M_*^{-1}$ is the size of the smallest possible semi-classical neutral non-rotating BH.     
   In four-dimensions, the bound then reads
  \cite{Dvali:2007hz, cesar, ramy}:\footnote{This
bound has been generalized for de Sitter spaces and cosmological backgrounds in \cite{Dvali:2008sy}.}
\begin{equation}\label{bhbound}
M_*\, = \, { M_4 \over \sqrt N_4}\, .
\end{equation}
Correspondingly in $4+d$-dimensions we have 
\begin{equation}\label{bhboundD}
M_*\, = \, {M_{4+d} \over  N_{4+d}^{{1 \over  2+d}}} \, , 
\end{equation}
where both $M_{4 + d}$  and $N_{4+d}$ must be understood from the point of view of $4+d$-dimensional theory.   That is,  $N_{4+d}$ counts the number of  $4+d$-dimensional species. 
Notice, that since the scale $M_*$ is unique for any given theory, the saturation of the bound 
in a high-dimensional theory automatically implies its saturation in a dimensionally-reduced 
theory.  This implies,  that the geometric relation between the Planck masses of high and low dimensional theories is controlled by the number of  KK species.   

 As said above, the scale $l_*$ marks the size below which BH can no longer maintain 
their semi-classical properties, such as,  the small relative rate of the temperature change, 
\begin{equation}
    {1 \over T^2} {dT \over dt}  \, \ll \, 1\, ,  
    \label{Tchange} 
 \end{equation} 
 where the temperature is the inverse of the gravitational radius $r_g \, = \, T^{-1}$. 
The latter relation must hold for any static, neutral and non-rotating semi-classical BH.   
The bounds (\ref{bhbound})  and (\ref{bhboundD}) follow from the fact that the semi-classicality condition (\ref{Tchange})  is violated for any BH smaller than $l_*$, due to an  unsustainably high rate of Hawking evaporation into the $N$ particle species \cite{Dvali:2007hz}.   

  Alternatively \cite{cesar}, the bound follows from the fact that at length scales shorter than $l_*$, the resolution of species is impossible {\it in principle}, since any detector capable of differentiating among the species at such a short length-scale would itself collapse into a BH.
  
 It is obvious, that the geometric relation (\ref{mplancks}) is a particular form  of (\ref{bhbound}). 
 For seeing this,  it is enough to notice that the factor $M_{4+d}^dV^{(d)} \, \equiv \, N_{KK}$ simply counts the number of KK species, whereas the high-dimensional Planck length is an obvious 
cutoff of the high-dimensional gravity theory. 
 
 As it follows from (\ref{bhbound}), in case the mass $M_*$ is of order TeV, one expects $N=10^{32}$
different species at or below this scale, and the strength of the gravitational interactions is diluted to its observed value due to the existence
of this large number of particles. Therefore, ADD model can be viewed as a particular representative of  
much more general class of the large species models
that address the hierarchy problem between the Planck scale and
the TeV scale.

   The purpose of the present work is to explore the physical meaning of  (\ref{bhbound}) and (\ref{bhboundD}) in string theory.  As explained above,  from the effective field theory perspective, string theory is a theory of infinite number of particle species.   Since, the bound relies on the BH evaporation, 
it is thus crucial to understand which are the relevant string species participating in this process. 
The question we would like to address now is, which particles play the role of the different species in D-brane string compactifications, and
how the black hole bound eqs.(\ref{bhbound}) and (\ref{bhboundD}) is satisfied in these string models.

In other words, we wish to understand how the string compactifications  fit into the large species scenario.  Thus, the following questions will be addressed:
\begin{itemize}
\item What is relation of the different mass scales in D-brane compactifications to the different particle species participating in BH evaporation?
\item How is the black hole bound satisfied by the stringy excitations?
\end{itemize}
As we shall see these questions can be relatively easily  answered for the KK modes, both for KK modes corresponding to the open strings on the
wrapped D-branes and also for the closed string KK modes in the bulk space, transversal to the D-branes.
However,  the most 
interesting and more subtle question
is, how the stringy excitation modes (Regge states) respect the bound (\ref{bhboundD}), because
naively the exponentially growing number of these states seems to violate it for any 
semi-classical BH smaller than the string length. 

  There are the two regimes that we shall explore.

  \subsection{Super-Stringy  Black Holes}
  
  The first regime explores the BHs  with the gravitational 
  radius larger than the string length  $r_g \, >  \, l_s$, where $l_s \equiv  M_s^{-1}$.   Such BHs can only evaporate into the lowest closed and open strings modes and their KK excitations. An obvious requirement, that such BHs must  be semi-classical,  implies the absolute upper bound on the value of the string scale
 \begin{equation}\label{boundzero}
M_s\, \leqslant \, {M_{4+d} \over N_{4+d}^{{1\over 2+d}} (m<M_s)}\, ,
\end{equation}
 where $N_{4+d}(m<M_s)$ counts the number of all $4+ d$-dimensional elementary particle species with masses below the string scale.  In a consistent compactification, the bound must be satisfied 
 for any $d$, and as we shall see, it indeed is. 

\subsection{Sub-Stringy Black Holes}

   The second regime, in which BHs horizon shrink  below the string length, $r_g \, < \,  l_s$, is much more profound.  The naive impression is, that such BHs simply cannot exist as semi-classical objects with the well-defined Hawking temperature, since for $T > M_s$ the number of thermally-accessible species 
   diverges exponentially and one is bound by the Hagedorn  effect.   The result would be an exponential increase of BH evaporation channels and the semi-classicality condition (\ref{Tchange})
   would be violated almost immediately.  
   We shall see, that this naive argument is false, and it alone cannot prevent the semi-classical BHs from
   continuing existence in sub-$l_s$ domain.  The reason is, that although number of states is indeed 
   exponentially large, only few of them are effectively produced in the BH evaporation.  
  Our findings indicate that the effective number of emitted string resonances is maximum
  \begin{equation}
  \label{neff}
  N_{eff} \, \sim \, 1/g_s^2 \, ,
  \end{equation}
 where $g_s$ is the string coupling constant\footnote{The alternative evidence \cite{cesar2} for this 
 bound comes from applying the arguments of \cite{cesar} that are restricting the information-storage by use of  species}.
  
    In order to illustrate this suppression, we shall develop an effective theory of BH evaporation into the string resonances.   The emission of a given string resonance is suppressed by a power of 
    $T/M_{4+d}$ defined by the oscillator number.  A crude intuitive understanding of this suppression can come from thinking of higher string resonances as of vibrating long strings.  
  Production of such a string by a tiny BH is only possible for certain patterns of the string vibration, namely if during its vibration the strings contracts down to a BH size.  For the other vibration patterns, the production  must be suppressed.    
  
    This simple qualitative picture is also supported by a toy model, which  
allows for a microscopic field-theoretic description of the BHs evaporation into stringy states. The latter  model is an $SU(n)$-QCD coupled to Einsteinian gravity.   The advantage of this system is, that it allows for the complementary descriptions of the micro BH evaporation both in the language of  closed strings (glueballs),  as well as in the language of elementary gluons.  The latter  description reproduces   (\ref{neff}). 

 Remarkably,  the relation (\ref{neff}) when substituted in (\ref{bhboundD}), reproduces the known relation between the string and the Planck scales
 \begin{equation}
  M_{10}^8 \, = \, {1 \over g_s^2} \, M_s^{8}  \, .
\label{msmp}
\end{equation} 
   Although this remarkable connection is indicative of intrinsic consistency of micro BH physics in string theory,   it has to be interpreted with care.  In particular, it cannot be regarded as a derivation of (\ref{msmp})  from the species bound (\ref{bhboundD}).   From the way we arrived to it, this relation by no means indicates that the 
 BH semiclassicality scale is bounded by $l_s$.    This would be the case if all $N_{eff}$ modes were opening up at or below the scale $M_s$.   However, they open up gradually at higher energies and 
 for $g_s \, \ll \, 1$ most of them are crowded  up at the temperatures $T \gg M_s$.  This is why, the BHs smaller than $l_s$ are not inconsistent with semiclassical BH physics. 
 
   Our results have some fundamental as well as phenomenological implications. 
 On one hand they indicate,  that  BH semi-classicality properties combined with the exponentially large number of string resonances  do not {\it a priory} prevent the small BHs from probing distances shorter 
 than the string length $l_s$.  In this respect,  micro BHs seem to share some properties of $D$-branes, 
 which are known to probe the scales parametrically shorter than the string length \cite{Douglas:1996yp}.  
 
   Secondly,  our findings shed some light at the expected properties of the micro BH that may be produced in particle collisions at super-stringy energies, e.g.,  in the high energy cosmic rays or at 
   LHC.    Since even in the most optimistic scenarios, the largest produced BHs at LHC will still be much smaller than the size of the extra dimensions, their evaporation will not be democratic in species
   \cite{Dvali:2008rm}. 
 Instead, such BHs  will predominantly evaporate into the species whose wave-function profiles in the extra dimensions have maximal overlaps with the BH localization site.  In this respect, democracy of the evaporation products will be a function of the BH size (and thus of the center of mass energy in 
 LHC collisions).  Observation of the mass-dependent democracy will be most easy within the Standard Model species, e.g.,  for the different quark and lepton  families,  since these species  cannot be too far displaced in the extra space, 
 due to their gauge quantum numbers.  For example, by gauge invariance, all the electrically-charged states must have the same overlap with the photon wave-function.  Thus, in this case, variation in
 BH democracy may be in principle noticeable even for a relatively small variations of the BH 
 mass.   For example,  heavier BHs produced in the collision of the first generation quarks, must evaporate more democratically into other flavors, than the lighter BHs produced in the same collision.  

  Finally,   after shrinking to the size $l_s$  BHs must evaporate into $N_{eff} \, \sim \, 1/g_s^2$ string resonances, some of which which may then be identified via their decay channels into the SM particles.

\section{Preliminaries} 
\subsection{The three fundamental length scales in brane compactifications:}

In the following we will consider type II orientifold compactifications with D-branes (for reviews see
\cite{Lust:2004ks,Blumenhagen:2006ci}).
Let us recall that
there are three basic, dimensional parameters in type II string compactifications with wrapped and/or intersecting D-branes.

\vskip0.2cm
\noindent
Of course,  first  there is the fundamental string scale, given in terms of the slope parameter $\alpha'$ as:
\begin{equation}
(1):\quad M_s={1\over \sqrt{\alpha'}}\, .
\end{equation}
The string scale is related to the fundamental $10$-dimensional Planck mass $M_{10}$ through 
a dimensionless string coupling constant $g_s$ via the relation (\ref{msmp})
 
 
Second,  we have to compactify from ten to four dimensions on an internal 6-dimensional space $\mathcal{M}_6$.  So as the second basic
string mass scale,  we introduce the scale defined by the  volume $V^{(6)}$ of the internal space:
\begin{equation}
(2):\quad  m_6 \, \equiv \, {1\over (V^{(6)})^{1/6}}\, .
\end{equation}
When measuring masses in terms of Planck units, it is
useful to introduce a dimensionless overall volume, given in $10$-dimensional Planck mass units, as
\begin{equation}
V_6' \, = \, V^{(6)}M_{10}^6 \, = \, {M_4^2\over g_s^{1/2} M_s^2}\, .
\end{equation}

We are considering scenarios, where the SM is located at some stacks of D(3+p)-branes, which are wrapped around 
some internal p-cycles inside $\mathcal{M}_6$:
$\Sigma_p\subset \mathcal{M}_6$. So, the third basic string mass scale is related to the volume $V_p^\parallel$ of the sub-space $\Sigma_p$, around which the D-branes
are wrapped:
\begin{equation}
(3):\quad m_p^\parallel={1\over (V_p^\parallel)^{1/p}}\, .
\end{equation}
Alternatively we can introduce the volume $V_{6-p}^\perp$ of the space transversal to the D-branes,
\begin{equation}
(3'):\quad m_{6-p}^\perp={1\over (V_{6-p}^\perp)^{1/(6-p)}}\, ,
\end{equation}
where the following relation holds:
\begin{equation}
V^{(6)}=V_p^\parallel V_{6-p}^\perp \, .
\end{equation}

\vskip0.2cm\noindent
These three fundamental dimensional parameters of D-brane models are linked to two 4D physical  observables in the following way:

\vskip0.2cm
\noindent
(A) the strength of gravitational interactions, as determined by the 4D Planck mass:
\begin{equation}\label{mplanck}
(A):\quad M_{4}^2= g_s^{-2}\ M_s^8\ V^{(6)}\ .
\end{equation}


\vskip0.2cm
\noindent
(B) the strength of the gauge interactions, as determined by the 4D gauge coupling:
\begin{equation}\label{Dpgaugecoupling}
(B):\quad g_{Dp}^{-2}= M_s^p\ g_s^{-1}\
V_p^\parallel\ .
\end{equation}
Hence, for fixed string coupling $g_s$, knowing $M_4$ and $g_{Dp}^{-2}$ leaves one of the three dimensionful string parameters
undetermined (unlike for the heterotic string). Since the 4D gauge coupling must not be too small, we take it of order unity.
This fixes $(V_p^\parallel)^{-1/p}$ to be of order $M_s$, i.e. $(V_p^\parallel)^{-1/p}\simeq M_s$. This leaves only $M_s$ (or $V^{(6)}$) as free parameter.
Then, the transversal volume $V_{6-p}^\perp$ is of the oder
\begin{equation}\label{transversal}
V_{6-p}^\perp=V^{(6)}M_s^p\, .
\end{equation}

\vskip0.2cm\noindent

From the model-building perspective there are several ``natural choices"  for $M_s$ 
which scan the mass scales starting from just below $M_{4} \sim 10^{19}$ GeV all the 
way to TeV scale. The later choice being motivated by the solution to the hierarchy problem, as well as 
by the prospect of the experimental discovery  of string theory  in particle physics and in  
table-top  experiments.

 \subsection{Choices of $M_s$ and low string scale compactifications}
  

  Although in our discussion $M_s$ will be kept as a free parameter,  we wish to briefly display
 some phenomenologically-motivated choices for it.

 We shall consider cases of $(V^{(6)})^{-1/6} \, \sim  \, M_s$ and $(V^{(6)})^{-1/6} \, \ll \, M_s$ separately. 
 The choice $(V^{(6)})^{-1/6} \, \sim  \, M_s$ is special, since in this case the super-$l_s$-size
 ($ r_g \, \gg \, l_s$) BHs evaporate only into the  zero modes of the lowest string excitations, and never in 
 their KK or Regge resonances. In the other words, for this choice the 
semi-classical BHs directly jump from four-dimensional Einsteinian regime 
to the stringy one in which they start evaporation into the Regge excitations.   
Notice, that in every other case, there is an intermediate regime, when 
 semi-classical BHs become high-dimensional and evaporate in KK resonances but not in the Regge ones.   

 
 
 \subsubsection{String Length Compactifications:  High String Scale}
  
  We shall first focus on the choice $(V^{(6)})^{-1/6} \, \sim  \, M_s$. Notice,  that for any given $g_s$ 
this choice  corresponds to the maximal proximity between the string and four-dimensional Planck masses. 
   This choice is also rather  special from the point of view 
of species counting.  As said above,  in this case  the larger than the $l_s$ BHs are automatically four-dimensional Einstein BHs, with the temperature $T \, < \, M_s, m_6$. Thus,  none of the  KK or Regge excitations are thermally accessible.   Thus,  the bound on species automatically translates as the bound on the possible number of the  four-dimensional zero modes.   We shall now express this bound as the relation between the number of zero modes and the string coupling.   

The relation between $M_s$ and $M_4$ 
 is given by
\begin{equation} 
M_s \, = \, {M_{4} \over g_s }\, ,  
\label{maximalMs}
\end{equation}
which follows from the relation (\ref{mplanck})   and (\ref{mplancks})
and from setting the compactification radius equal to $l_s$.
 
  The realistic numerical value of $M_s$  can be estimated by using (\ref{Dpgaugecoupling}) and assuming the standard UV boundary condition for the running gauge coupling $g_{Dp}$.    We thus get  the estimate for the highest phenomenologically-consistent choice of $M_s$ in the weakly-coupled description (which is similar to the value in  heterotic string compactifications): 
\begin{equation}
M_s\, \sim  \, 10^{17-18} ~{\rm GeV} \, .
\end{equation}
The masses of  all the KK excitations as well as of the string Regge  modes are at or above $ M_s$ and are irrelevant for the dynamics of semi-classical BHs of size $r_g \, \gg \,  l_s$.   Since
such BHs are larger than both the string length as well as the the compactification radius, they are 
well within the domain of semiclassical description and for all practical purposes represent the  four-dimensional  Schwarzschild  BHs.   Their semi-classicality  immediately  implies the bound 
(\ref{bhbound}), which dictates that the number of all possible zero modes in such compactifications
is automatically restricted by 
\begin{equation}
  N_{zero} \, \lesssim \, {1 \over g_s^2} \, .
  \label{boundzero}
  \end{equation} 

 This bound is absolute, since, by default,  at distances $\gg l_s$ there are no new gravitational modes available that could modify the  universal thermal properties  of the Einsteinian BHs. 
 The only gravitational degree of freedom at large distances is a four-dimensional massless graviton. Because of this,  the classical metric of neutral BHs, by  standard no-hair theorems \cite{nohair}, is exactly
 Schwarzschild.   Thus, the energy loss of such a BH is 
  \begin{equation}
  {d M_{BH} \over dt}  \,  =  \, - \, T^2 \, N_{zero} \, ,
  \label{zeroloss}
  \end{equation}
which implies the breakdown of the semi-classicality condition (\ref{Tchange}) at the  temperature 
\begin{equation}
T_*  \, = \ M_* \, = \, {M_4 \over \sqrt{N_{zero}}}\, .
\label{tstar}
\end{equation}
Since by consistency $T_*$ cannot be below $M_s$ (below string scale there are no new modes available for changing the semi-classical regime),  taking into the account  (\ref{maximalMs}), we prove  (\ref{boundzero}).

\subsubsection{Large Volume Compactifications: Low String Scale} 

    Choices of $M_s$ with values lower than (\ref{maximalMs}),  imply 
   at least some compactification radii larger than the string length  $l_s$.  In this case the evaporating semi-classical BHs cross over to a high-dimensional regime before reaching the 
   string size, and  evaporate not only into the zero modes
   but also in their KK excitations.  The evaporation of such BHs will be considered in details in the next section. 
   Here we just list the few interesting choices of  $M_s$ values.  
   
Motivated by gauge coupling unification,  one often assumes
the existence of a GUT gauge group and makes the identification:
\begin{equation}
M_s\equiv M_{GUT}\simeq 10^{16}~{\rm GeV}\, .
\end{equation}
 In this scenario, $V^{(6)}$ is of order 
 \begin{equation}(V^{(6)})^{1/6}\simeq 10^{-15}~{\rm GeV}^{-1}\, .
 \end{equation}
 
 Another choice of lower string scale compactifications is to identify $M_s$ with the scale of 
spontaneous supersymmetry breaking:
\begin{equation}
M_s\equiv M_{SUSY}\simeq 10^{11}~{\rm GeV}\, .
\end{equation}
 The rational for this choice is that one does not have to worry about  lowering the scale 
 of supersymmetry breaking below the cutoff, since the gravity-mediated contribution to the 
 Standard Model masses are automatically of the right order of magnitude.   However, 
 one has to be careful in avoiding $1/M_s$-suppressed interaction between the Standard Model 
 and the supersymmetry-breaking  sectors,  which would destabilize the hierarchy.  
 
 This scenario implies that $V^{(6)}$ is of order 
\begin{equation}
(V^{(6)})^{1/6}\simeq 10^{-(6-7)}~{\rm  GeV}^{-1}\, .
\end{equation}

The most radical choice (motivated by the hierarchy problem) 
is to identify $M_s$ right away  with the scale of the SM,
i.e. with the TeV scale:
\begin{equation}
M_s\equiv M_{SM}\simeq 10^{3}~{\rm GeV}\, .
\end{equation}
Therefore,  in this scenario, $V^{(6)}$ is very large, and is of order 
\begin{equation}
(V^{(6)})^{1/6}\simeq 10^{14/6}~ {\rm GeV}^{-1}\, , 
\end{equation}
and hence this scenario
can also be called {\sl very large extra dimension scenario.}

\vskip0.3cm
\noindent
Let us summarize these four possibilities, and the corresponding energy and length scales in the  table 1 (we use that
$1~{\rm GeV}\sim (10^{-16}~{\rm m})^{-1}$, we have set $g_s\simeq 1$, $p=4$, and we neglect factors
of $2\pi$).
$V_6'$ takes the following values:
\begin{equation}
V_6'\, \sim \, 1,10^6,10^{16},10^{32}\, .
\end{equation}

\begin{table}\scriptsize
\caption{The different mass scales in D-brane models}
\label{Table1}
\centering
\vspace{3mm}
\begin{tabular}{|c||c|c|c|c|c|c|}
\hline
 & $M_s$ (GeV) & $L_s$ (m) & $m_6=(V^{(6)})^{-1/6}$ (GeV)& $(V^{(6)})^{1/6}$ (m) & $m_2^\perp=(V_2^\perp)^{-1/2}$ (GeV) & $(V_2^\perp)^{1/2}$ (m)\\
 \hline
 \hline
 (o) & $10^{18}$ &$10^{-35}$ & $10^{18}$ & $10^{-35}$ & $10^{18}$ & $10^{-35}$\\
  \hline
 (i) & $10^{16}$ &$10^{-32}$ & $10^{15}$ & $10^{-31}$ & $10^{13}$ & $10^{-29}$\\
  \hline
 (ii) & $10^{11}$ &$10^{-27}$ & $10^{6-7}$ & $10^{-(22-23)}$ & $10^{3}$ & $10^{-19}$\\
 \hline
 (iii) & $10^{3}$ &$10^{-19}$ & $10^{-14/6}$ & $10^{-14}$ & $10^{-13}$ & $10^{-3}$\\
 \hline
\hline
\end{tabular}
\end{table}

\subsection{Dynamical realization of low string scale compactifications}

 In our analysis,  we shall make no specific assumption on how the moduli are stabilized. Our results will apply to any consistent static or a nearly static compactification.  
   What is important is, that the possible time-dependence of the background 
  is slower than the evaporation process of the {\it  smallest}  semi-classical BHs existing on the background of interest
\cite{Dvali:2008sy}.  
  
   For example, for the bounds  obtained  via  the four-dimensional semi-classical BHs this will imply, that there must exist  such BHs for  which 
 the  evaporation time is shorter than the effective four-dimensional Hubble time.  Needless to say,  for any realistic compactification this condition must be automatically satisfied.  
   
    As long as the above condition is satisfied, the precise mechanism of  compactification and 
 moduli stabilization is unimportant for our analysis.   On the backgrounds of our interest, the moduli fields may be:   Stable (positive masses-squares), neutrally-static (massless)  or  even slowly evolving in time (either tachyonic or runaway) . 
 As long as the background supports the existence of semi-classical BHs, which evaporate faster than the characteristic  time of the background-evolution, our bounds are applicable.  
 
  As for the high-dimensional or sub-$l_s$ BHs, the issue of compactification plays no role,  since their evaporation can be considered directly in ten-dimensional theory.  As long as throughout their half-life they stay smaller than the size (or the curvature radius) of the extra space, the time-dependence of the background is unimportant.   
   
     We thus see, that derivation of our bounds is rather insensitive to the stabilization of the radii of the compactified space.  The converse, however, is not true. The BH bound on number of species puts severe 
constraints on the possible static (or slowly-changing) backgrounds in string compactifications. 
For example, any background that violates (\ref{boundzero}),  must be {\it clasically-unstable} with the instability time less than the lifetime of any  Schwarszchild  BH of  size larger than the compactification radius.

\vskip0.5cm
\noindent

Of course,  as the necessary  model-building aspect, one  has to address the issue of  how moduli stabilization might lead to such big internal volumes, as required in
scenarios (ii) and (iii).   This issue will not concern our work.  However, our results can be used as an useful consistency checks of such compactifications, in terms of number of species.   

 Some plausible mechanisms  in the direction of moduli stabilization in large volume compactification  were  developed in \cite{Balasubramanian:2005zx}, where in was argued that  for the case of the SUSY breaking scenario (ii), loop effect and non perturbative effects
in the effective moduli potential can easily and naturally lead to such big, exponential hierarchies.
The relevant geometries are called Swiss cheese Calabi-Yau space, where the D-branes of SM model are wrapped
around small cycles (holes) inside a huge internal space.

In order to be a little bit more specific, let us recall the main points in the construction of large volume D-brane
comapctifications. The effective field theory potential ${\cal V}$ for the moduli fields is based on the following 
K\"ahler potential and superpotential   \cite{Balasubramanian:2005zx,Conlon:2005ki}
\begin{eqnarray}
{{\cal K}\over M_4^2}&=&-2\log\biggl(V_6'+{\xi \over 2}\biggr)-\log\biggl( -i(\tau-\bar \tau)\biggr)+{\cal K}_{cs}\, ,
\nonumber\\
W&=&{g_s^{3\over 2}M_4^3\over \sqrt{4\pi}}\sum_i A_ie^{-{2\pi T_i\over g_s}}+W_{cs}\, ,
\end{eqnarray}
where $\tau$ denotes the axion-dilaton field, the $T_i$ correspond to the volumes of the 4-cycles, being wrapped by D7-branes,
the $\xi$-terms is a higher order loop correction to the K\"ahler potential, and ${\cal K}_{cs}$, $W_{cs}$ are functions, which depend
on the complex structure moduli fields. Using ${\cal K}$ and $W$ it is straightforward to compute the standard supergravity scalar potential
given as
\begin{equation}
{\cal V}=e^{\cal K}\biggl( |DW|^2-3|W|^2\biggr)\, .
\end{equation}
As it was shown in \cite{Balasubramanian:2005zx,Conlon:2005ki} it is possible to obtain minima of ${\cal V}$ with hierarchically large
values for $V_6'\sim 10^{16}$ and correspondingly low values of the string scale $M_s$. This depends crucially on the loop correction $\xi$
to the K\"ahler potential as well as on the exponential dependence of $W$ on the moduli $T_i$, as it was demonstrated in 
\cite{Balasubramanian:2005zx,Conlon:2005ki} for the SUSY breaking scenario (ii). However it is equally possible by a small variation
of the parameters in ${\cal K}$ and $W$ to obtain minima of ${\cal V}$, where $V_6'$ takes even much larger values of the order
$10^{32}$.\footnote{We like to thank F. Quevedo for discussion on this point.}
However,  the masses of 
some moduli fields obtained from this scalar potentiual can be unacceptably small in case the volume $V_6'$ is very large
(see \cite{Conlon:2005ki} for a table with masses computed from this potential).
We will discuss this potential problem in the next section for the overall modulus field $T$, and we will argue
that generic radiative mass corrections to the overall $T$-field modulus will lift its mass to a phenomenologically acceptable value.

\section{Several generic particles in brane compactifications and their associated length scales}

Now let us recall several generic particles that are related to the different mass scales in D-brane
compactifications, as well as their masses, measured in the Einstein frame, and
expressed in terms of Planck units.

\vskip0.3cm
\noindent
{\it (i) String Regge excitations:}

\vskip0.2cm
\noindent
The most generic and model independent stringy excitations are the so-called Regge 
modes. Their masses are proportional to the string scale $M_s$, and in terms of Planck units given by:
\begin{equation}
M_{\rm Regge}=\sqrt nM_s={\sqrt n M_4 \over g_s^{1/4}\sqrt{V_6'}}\, .
\end{equation}
Here $n$ denotes the oscillator number of these states. In case they are open strings, 
the stringy Regge excitations can carry the SM quantum numbers, i.e. they can be produced by pp-collisions via gauge interaction. 
Their discovery in LHC experiments is then possible in the TeV scale string scenario \cite{Lust:2008qc,Anchordoqui:2008di}.

\vskip0.3cm
\noindent
{\it (ii) Transversal direction KK particles:}

\vskip0.2cm
\noindent
For the masses of the KK particles related to the large volume cycles  one obtains:
\begin{equation}
M_{KK}^\perp={m \over (V_{6-p}^\perp)^{1/(6-p)}}\, = \, m~M_{4} (V_6')^{{8-p\over 2p-12}}\, ,
\end{equation}
where $m$ counts the internal closed string KK momentum of each state.
In the four scenarios discussed above $M_{KK}^\perp$ takes the following values (here for $m=1$, $p=4$):
\begin{equation}
M_{KK}^\perp\, \sim \, 10^{18},10^{13},10^{3},10^{-13}~{\rm GeV}\, .
\end{equation}
These KK particles are closed strings, and they are neutral under the SM gauge group, and hence interact only gravitationally with SM particles.

\vskip0.3cm
\noindent
{\it (iii) Small (longitudinal) cycle KK particles:}

\vskip0.2cm
\noindent

For the masses of the KK particles related to the small volume cycles  one obtains:
\begin{equation}
M_{KK}^\parallel={m' \over (V_p^\parallel)^{1/p}}
\end{equation}
Here $m'$ is the open string KK momentum along the wrapped $(3+p)$-branes.
For the compactification on the parallel directions being $l_s$,
in the four scenarios discussed above $M_{KK}^\parallel$ takes the following values ($m'=1$):
\begin{equation}
M_{KK}^\parallel=10^{19},10^{16},10^{11},10^{3}~{\rm GeV}\, .
\end{equation}
These KK particles are open strings, and hence they carry SM quantum numbers, i.e. they are the KK excitations of the SM fields. Hence they can be
directly produced at the LHC.

\vskip0.3cm
\noindent
{\it (iv) Overall volume modulus:}

\vskip0.2cm
\noindent
Related to the overall six-dimensional volume $V_6$, there is a generic closed string modulus field $T$ 
in any compactification.  In the low string scale compactifications, the tree-level  mass of this modulus 
may be unacceptably small,   
(e.g.,  its typical tree level mass was calculated using the moduli potential calculated in \cite{Conlon:2005ki}: 
$M_T={M_{\rm Planck}\over (V_6')^{3/2}}$). 

However, the point we would like to stress is, that there is a model-independent lower bound on 
the typical mass that the volume modulus should get from the quantum correction, which is given by 
\begin{equation}
\Delta M_T^2 \, \sim  \,  {\langle T_\mu^\mu T_\mu^\mu \rangle \over M_{4}^2} \gtrsim \, {\Lambda_{SM}^4\over M_{4}^2}\,,   
\label{Tbound}
\end{equation} 
where  $\Lambda_{SM}$ is the scale that cuts-off the contribution to the $\langle T_\mu^\mu T_\mu^\mu \rangle$ correlator  from the loops of the standard model particles.   The source of this contribution 
is in the fact that the volume modulus is coupled to trace of the energy-momentum tensor, 
$T_{\mu}^{\mu}$,   gravitationally by (at most) 
$1/M_4$-suppressed interactions.   Thus,  the Standard Model loops inevitably contribute to the 
generation of its mass.  The relevant diagrams are loops with the external $T$-legs, and with the Standard Model particles circulating in the loop. Notice, that $\Lambda_{SM}$  is entirely determined  by the scale at which the momenta of the Standard Model particles flowing in the loop  get regulated, and is  independent of the 
momenta flowing in the external $T$-legs.   Due to this, the result is insensitive to the UV properties of the effective coupling between the standard Model species and the volume modulus, which makes 
(\ref{Tbound}) very robust.  An obvious  lower experimental 
bound on $\Lambda_{SM}$ is around TeV, which taking into the account  the multiplicity of the contributing species (and ignoring the accidental cancellations)  translates as the expected lowest value of the volume modulus mass around  
\begin{equation}
\Delta M_T \, \sim \, 10^{-3} ~{\rm eV}\, ,
\label{Tlowest}
\end{equation}
which corresponds to a length scale in sub-millimeter range. This is the  range where the existing 
table-top fifth force experiments \cite{adelberger}  start to penetrate. The resulting gravity-competing force should share similarity with the ones considered previously from light moduli  in gauge-mediated supersymmetric theories \cite{savasgian} and from volume-modulus in orbifolds string compactifications with Scherk-Schwarz supersymmetry breaking \cite{submmforce}.    
 
Notice,  however, that  in many cases,  the expected mass in reality  must be somewhat higher, either due to 
multiple sectors contributing in the $\langle T_\mu^\mu T_\mu^\mu \rangle$-correlator, or due to higher cutoff  (or both).    


\vskip0.3cm
\noindent
{\it (v) Small (longitudinal) cycle volume moduli:}

\vskip0.2cm
\noindent
Related to the small cycles of longitudinal volume $V_p^\parallel$, there are a moduli fields $t$ with tree level masses (here for $p=4$)
\begin{equation}
M_t={M_{4}\over V_6'} \, .
\end{equation}
In the four scenarios discussed above $M_t$ takes the following values:
\begin{equation}
M_t=10^{19},10^{13},10^{3},10^{-13}~{\rm GeV}\, .
\end{equation}
Again, these t-moduli are neutral and interact only gravitationally with SM fields.



\section{Black Hole evaporation in String D-brane compactifications and the bound on species-number}

We now wish to discuss the question, in what sense string D-brane comapctifiactions are a concrete realization of the large species-number scenario of \cite{Dvali:2007hz}. Namely we want to analyze how the different particles species in D-brane
models respect the BH bound eq.(\ref{bhbound}). 

  Since the number entering in (\ref{bhbound}) is the effective number of species to which semi-classical 
BHs can evaporate, we thus focus our analysis on the study of micro-BH evaporation process in string compactifications.    We shall consider the regimes $r_g \, \gg \, l_s$ and $r_g \, < \, l_s$ separately. 
 In the first case the string  Regge states are unaccessible and the only relevant degrees of freedom are 
 zero modes and their KK  excitations.   We start with this case first.

\subsection{Evaporation into KK particles}

 In this section we shall work  under the choice $g_s \, = \, 1$. This choice allows us to avoid possible production of the higher string Regge 
excitations in any semi-classical BH evaporation, since the latter by default must have a temperature 
below $M_{10}$.   The particles contributing in the evaporation of such BHs are  
only the string zero modes and their KK resonances.  Hence, these will be the states contributing 
into the BH bounds (\ref{bhbound}) and (\ref{bhboundD}), and our goal in this section will be 
to observe  how the consistency of these bounds is maintained. 

 We shall  consider sub-$l_s$ BH regime in the next section, where we shall allow for the hierarchy 
 between the  $M_s$  and $M_{10}$ scales.  

     In any $4+d$-dimensional scenario, with the coordinates in extra $d$-dimensional space denoted by $y_{1}, ...y_{d}$, the BH evaporation rate  is given by 
     \begin{equation}
      {d M_{BH} \over dt}  \, = \, - \, T^2 \, N_{eff} \, ,
    \label{BHDrate} 
\end{equation}
where,  it is assumed that the BH is a perfect quantum emitter,  with the temperature given by 
$T\, = \, r_g^{-1}$.   The effective number of particle species $N_{eff}$ is given by
\begin{equation} 
N_{eff} \, \simeq \, \sum_{m}  \, {\rm e}^{-{m\over T}}\,  \int_{|y| \, < \, r_g} \,  |\psi^{(m)}(y)|^2 \, \, 
\label{neffD}
\end{equation}
where the exponent stands for the Boltzmann suppression, and  the $d$-dimensional integral 
evaluates the overlap of each particle wave-function profile with the BH horizon.  

 To be most precise, the integral has to be taken over a background metric in the presence of a {\it classically-static} BH. However, for the micro BHs with short evaporation time, this subtlety can be ignored,  and in the leading approximation the deformations of the particle wave-function profiles due to the BH metric can be disregarded. 
  
 Of course, for the  BHs with the size $r_g$ smaller than the size of an extra dimension, the corresponding high-dimensional relation between the mass and the Hawking temperature must be assumed. 

 The key point then is,  that  the bound (\ref{bhboundD}) between the gravitational cutoff of the theory 
 $M_*$ and $M_{4+d}$  is reproduced by $N_{d+4} \, = \, N_{eff}$.  Moreover, since the scale $M_*$ is the universal cutoff,  the saturation of the bound 
 in one particular dimensionality, automatically saturates it in all.   
E.g.,  the four-dimensional bound (\ref{bhbound}) is  simultaneously reproduced by $N_{4}$ being the total number of four-dimensional species, participating in the sum in (\ref{neffD}). 
 
   We shall now illustrate this  point on some examples.




\vskip0.3cm
\noindent
{\it (i) Isotropic compact space}
\vskip0.2cm
\noindent
We first discuss the simplest case of an isotropic, d-dimensional compact space of volume $V_d=R^d$. (This includes
the case of  no (wrapped)  branes) Then the masses $M_{KK}^i$ of the bulk  KK-particles are given as
\begin{equation}
M_{KK}^i={\sqrt{\sum_i^dm_i^2}\over R}\, ,\quad (i=1,\dots ,d)
\, ,
\end{equation}
where the $m_i$ are the KK quantum numbers in the $i^{th.}$ direction of the internal space.
From above it follows that the total number of  KK particle with the masses  up to the scale $M_{4+d}$, 
for each $4+d$-dimensional species  is given by
\begin{equation}
N_{KK}=(M_{4+d}R)^d\, .
\end{equation}
Applying this counting to a theory with a single $4+d$-dimensional graviton, and using (\ref{bhbound}) with identification $M_* \, = \, M_{4+d}$,  we obtain the expression for the  effective Planck scale in four dimensions in terms of number of KK-species, 
\begin{equation}
M_4^2=M_{4+d}^2N_{KK} \, = \, M_{4+d}^2 \, (M_{4+d} R)^d\, ,
\label{M4KK} 
\end{equation}
which indeed agrees with the geometric expression for the Planck mass in four dimensions after compactification \cite{ArkaniHamed:1998rs}.

 On the other hand the BH bound (\ref{bhbound}) counts the total number $N$ of four-dimensional species, which in case of $N_{4+d}$  high-dimensional species increases as  $N \, = \,  N_{4+d}  \, N_{KK}$. 
Substituting this in (\ref{bhbound}) and taking into the account (\ref{M4KK}),  we automatically get 
(\ref{bhboundD}). 

For $N_{4+d} \, \sim  \, 1$, which is the case for example for pure gravity in $4+d$-dimensions, 
 the both BH bounds,  (\ref{bhbound}) and (\ref{bhboundD}), are automatically satisfied  by $M_* \, = \, M_{4+d}$, as long as $M_{KK}\, < \, M_{4+d}$ or equivalently as long as  
\begin{equation}
R>1/M_{4+d}\, .
\end{equation}

However,  for $N_{4+d} \, \gg \, 1$, the bound indicates that the true gravitational cutoff, must be below 
$M_{4+d}$!
This result is absolutely compatible with our understanding of the structure of string theory in ten-dimensions. 
Indeed,  for $g_s \sim 1$,   the number of ten-dimensional species with mass below $M_{10}$
 is limited by the fixed number of closed string zero modes, 
and cannot be arbitrarily large.  That is,  for $g_s \sim 1$,  $N_{10} \, = \,  N_{zero}$.  
   Only possibility for increasing $N_{10}$  arbitrarily is to take a weak string coupling 
   $g_s \rightarrow  0$, but then the string scale $M_s$, which is the gravitational cutoff of the theory,  
inevitably comes below $M_{10}$.  

 Let us now check how the above mode-counting reproduces the correct relation between the $M_{4+d}$ and the 
 $M_*$ through the relation (\ref{BHDrate}).   For a BH of size $r_g \, \ll  \, R$,  the relation between the mass and the temperature is, 
 \begin{equation}
 M_{BH} \, = \, M_{4+d} (M_{4+d} T^{-1})^{1+d} \, .
 \label{MBHandT}
 \end{equation}
Noticing,  that the overlap between the KK wave-functions and the BH horizon is 
$(r_g/R)^d$, and the thermally available number of KK states per each $4+d$-dimensional particle  is $N_{KK}(m<T) \, = \, (TR)^d$, the value of $N_{eff}$ in (\ref{neffD}) becomes,
\begin{equation}
N_{eff} \, = \, (r_g/R)^d (TR)^d N_{4+d} \, = \, N_{4+d} \, .
\label{neffnew}
\end{equation}
Using (\ref{MBHandT}) and (\ref{neffnew}), the  eq(\ref{BHDrate}) can be written as, 
    \begin{equation}
     {1 \over T^2}\,  {dT \over dt}  \, =  {1 \over (1+d)} \, \, \left ( {T \over M_{4+d}} \right)^{2+d} \, N_{4+d}  \, .     
    \label{TchangeD} 
\end{equation}
 It is obvious that the semi-classicality condition (\ref{Tchange}) breaks down exactly at $T \, = \, M_*$, where  $M_*$ is given by (\ref{bhboundD}).   There is a complete consistency between the BH bounds obtained in four and high-dimensional theories.

\vskip0.3cm
\noindent
{\it (ii) Non-isotropic compact space -- wrapped D-branes}
\vskip0.2cm
\noindent
Now we are ready to  discuss the case of non-isotropic extra space,  in which we have $(3+p)$-branes wrapped around internal p-cycles.
The longitudinal volume of the wrapped branes is $V_{\parallel} \, = \,R_{\parallel}^p$, whereas the transversal volume
is $V_{\perp}\, = \, R_{\perp}^{d-p}$. Hence the total volume is $V_d \, = \, V_{\parallel}V_{\perp} \, = \, R_{\parallel}^pR_{\perp}^{d-p}$.
As before, the total number of KK particles produced by  each $d+4$-dimensional degree of freedom  is given by
\begin{equation}
N_{KK,total} \, = \, (M_{4+d})^d \, R_{\parallel}^pR_{\perp}^{d-p}\, ,
\label{KKtotal}
\end{equation}
and hence if we had a single $4+d$-dimensional particle, we would again reproduce  
the geometric relation for $M_4$ in terms of number of species \cite{Dvali:2007hz}, 
\begin{equation}
M_4^2 \, = \, M_{4+d}^2N_{KK,total} \, = \, (M_{4+d})^{d+2}V_d\, .
\end{equation}
Again $M_4$ coincides with the Planck mass eq.(\ref{mplanck}) in string compactifications with $d=6$.

\vskip0.3cm

However, we now  have two different kinds of KK species. The first kind are the KK excitations along the wrapped p-branes.
Their masses and numbers are given as:
\begin{equation}
M_{KK}^\parallel={\sqrt{\sum_{i=1}^p(m'_i)^2} \over R_\parallel}\, ,\quad N_{KK,\parallel}=(M_{4+d}R_\parallel)^p\, .
\end{equation}

\vskip0.3cm

The second type are  the KK excitations in the bulk directions, transversal to the wrapped p-branes. Their masses and numbers are given as:
\begin{equation}
M_{KK}^\perp={\sqrt{\sum_{i=p+1}^dm_i^2} \over 
R_{\perp}}\, ,\quad N_{KK,\perp}=(M_{4+d}R_\perp)^{d-p}\, .
\end{equation}
Note that $N_{KK,total}=N_{KK,\parallel}N_{KK,\perp}$.

  Thus,  we have to distinguish two types of 
 high-dimensional particles as seen by a $4+d$-dimensional observer.   The first  category are
 $4+d$ dimensional bulk species of number $N_{4+d}$,  which poses KK excitations both in transverse as well as in longitudinal direction.   The total number of the  KK-excitations in both directions is given by  (\ref{KKtotal}).   The second category are the $4+p$-dimensional particles localized on the
 $p$-brane world volume.   These particles have no transverse momentum and thus only poses KK excitations in longitudinal directions.  Each of these localized states  deposits $N_{KK,\parallel}$  four-dimensional KK species.   
 
   We are now ready to generalize the BH evaporation analysis to the unisotropic  situation.
   One small complication is, that the BH evaporation rate now depends on its  relative location 
  with respect to the $p$-brane.  For instance,  a small  BH that is positioned away from the 
    brane cannot evaporate into the $4+p$-dimensional modes that  are localized on the brane. 
   It was suggested  in \cite{Dvali:2008rm}, that such BHs that are decoupled from the brane species 
   cannot be {\it classically-static}, and must  evolve in  time until they enclose the brane. 
  However,  if the BH is small enough and the initial distance to the  brane it sufficiently large, the  time scale of classical evolution can be much longer than the   evaporation time.  In this case the BH shall evaporate practically only into the bulk species. 
   In order to avoid irrelevant details of the above  complications, we shall assume that the BH is pierced by the brane to start with, and thus all the brane-localized species are readily accessible.   
  
   The BH in question will undergo the different evaporation regimes depending on its size (and thus temperature). 
    The  BHs that are  larger than the size of longitudinal dimensions, but still smaller than the transverse ones  $R_{\parallel} \, \ll \,  r_g \, \ll \, R_{\perp}$, will evaporate  as effectively $(4+d-p)$-dimensional. 
   The analysis of the previous section can be directly applied to such a BH via a simple  replacement  
      $d \rightarrow d-p$.  The only caveat is, that  $N_{eff}$ will also include $N_{4+p}$ zero modes 
   localized on the $p$-brane to which the BH will evaporate without any extra suppression. 
    In this respect,  each zero mode localized on the $p$-brane counts as a whole transverse KK tower of a single $4+d$-dimensional bulk state.  
       
     Let us now turn to the evaporation of a small BH,  $r_g \, \ll \, R_{\parallel}$. 
     The  mass-to-temperature relation of such a BH is the same as (\ref{MBHandT}). The only novelty in comparison with the previous section is the anisotropy of the evaporation into the transverse and the longitudinal  KK modes.  The production rate of the individual longitudinal and transverse KK modes  are  
suppressed by the wave-function overlap suppression factors, equal to  $(r_g/R_{\parallel})^p$ and 
     $(r_g/R_{\perp})^{d-p}$ respectively. However, these suppression factors are exactly compensated by the corresponding numbers of the thermally available 
     states, which are given by  $N_{KK\parallel} (m<T) \, = \, (TR_{\parallel})^p$ and $N_{KK\perp} (m<T) \, = \, (TR_{\perp})^{d - p}$  respectively. 
       As a result, every $4+d$-dimensional bulk degree of freedom as well as every $4+p$-dimensional brane mode,  each contribute as one mode into $N_{eff}$.  That is, 
       \begin{equation}
N_{eff} \, = \, N_{4+d}\, N_{4+p} \, , 
\label{neffasym}
\end{equation}
 and 
 \begin{equation}
   {1 \over T^2}\,  {dT \over dt}  \, =  {1 \over (1+d)} \, \left ( {T \over M_{4+d}} \right)^{2+d} \, N_{4+d}  \, N_{4+p} \, .     
    \label{dTasym} 
\end{equation}
 The  breakdown of the semi-classicality condition (\ref{Tchange})  again happens at the scale $M_*$ given by (\ref{bhboundD}) where $N_{4+d}$ has to be replaces by  $N_{4+d}  \, N_{4+p} $. 
  Thus, in creating the hierarchy between the Planck mass and the gravitational cutoff,  bulk and brane modes contribute equally.  Again,  we see a complete consistency between the four- and  the high-dimensional BH bounds. 
  
    Notice,  that the relation (\ref{neffasym}) tells us that every brane mode has the same effect on decreasing gravitational cutoff of the theory as the bulk mode.  So one could  reduce $M_*$ relative to
    $M_{4+d}$ at the expense of increasing $N_{4+p}$ rather than $N_{4+d}$. 
  However,  as was shown in  \cite{Dvali:2009fv},  the limit 
    $N_{4+p} \, \gg \, N_{4+d}$ is very subtle, since in this limit the gravitational effects of the 
    $p$-branes can  no loner be ignored. For instance,  in case of many coincident branes, they develop a horizon, and the BH evaporation into the brane modes changes at shorter scales.

\subsection{{Black hole evaporation into the string Regge excitations -- the effect of the string tower}}

It is known that in string theory the density of states grows exponentially for $M \, > \, M_s$:
\begin{equation}\label{expnumber}
\rho(M)  \, \sim e^{\sqrt{bd} {M\over M_s}} \, ,
\end{equation}
where $M_s$ is the string mass scale,  $b$ is a numerical constant and  the factor $d$ accounts for the fact that in $D$ space dimensions each string oscillator can come in $d =  D-2$ varieties. 

The obvious question is what is the implication of this exponentially-growing number  for the 
BH bound (\ref{bhboundD}). 

 A simple quick answer to the above question would be, that, since $M_s$ is an obvious cutoff, whereas all the higher Regge states are heavier,   they simply should not be counted 
 in eq.(\ref{bhboundD}).  Hence the only relevant species  would be the string zero modes and their KK excitations, which we have already taken into the account in the previous section. 

  This answer, however, is far from being satisfactory.  The reason is,  that the number $N_{4+d}$ 
  is the number of species to which the semi-classical BH can evaporate.  Can semi-classical 
  BHs smaller than $l_s$ exist?  Naively,  the answer to this question is negative, as the exponentially growing numbers seems to be in conflict with the black hole semi-classicality 
for temperatures above  $T > M_s$, similarly to the usual Hagedorn phenomenon.   Indeed, for the exponentially-growing $N$,  the condition (\ref{Tchange}) would be violated almost immediately.   We shall argue however, that this naive view is not supported by more detailed analysis. 
In fact, we shall discover quite the opposite,  that  the decay rate of  small BHs  into the stringy Regge states is suppressed.
In other words, the effective number of string species, into which the the black hole can decay, is smaller
that the exponentially growing number (\ref{expnumber}).   Thus, the thermal arguments do not 
forbid existence of the substringy size BHs,  and the BH  bound
is not violated by the string Regge exciaitions. 

Before we start the discussion about the black hole decays, let us note that
the above-mentioned suppression of the string Regge modes in the BH evaporation  is qualitatively similar to the rate of scattering of light SM particles into string Regge excitations, which exhibits Veneziano-type softening. 
In other words, in the perturbative string regime the excitation  of  the Regge states is effectively suppressed in scattering proceses
among the light fields.
The relevant scattering amplitudes
were recently computed in \cite{Lust:2008qc}.
As an example consider  the $2\rightarrow 2$ scattering amplitudes of 4 open string gluons on the D-brane, or 2 gluons and 2 quarks
with an intermediate Regge particles exchanged. This amplitude is only dominated by the exchange of string Regge excitations, no
KK particles can be exchanged in this process (KK particles are exchanged in 4-point scattering of 4 quark fields).
Generically, this 4-point, tree--level string amplitudes 
are described by the Euler Beta function depending on the kinematic
invariants $s=-(k_1+k_2)^2,\ t=-(k_1+k_3)^2,\ u=-(k_1+k_4)^2$, with $s+t+u=0$ and
$k_i$ the four external momenta.
The whole amplitudes $A(k_1,k_2,k_3,k_4;\alpha')$ may be understood as an  
infinite sum over $s$--channel poles with intermediate string Regge states
$|k;n\rangle$ exchanged.
After neglecting kinematical factors the string amplitude $A(k_1,k_2,k_3,k_4;\alpha')$ takes the form 
\begin{equation}\label{basic}
A(k_1,k_2,k_3,k_4;\alpha')
\sim-\frac{\Gamma(-\alpha' s)\ \Gamma(1-\alpha' u)}{\Gamma(-\alpha' s-\alpha' u)}
=\sum_{n=0}^\infty\ \ \frac{\gamma(n)}{s-M_n^2}
\end{equation}
as an infinite sum over $s$--channel poles at the masses
\begin{equation}
M_n^2=M_s^2\ n
\end{equation}
of the string Regge excitations.
In (\ref{basic}) the factor $\gamma(n)$  is the three--point coupling of the intermediate 
states $|k;n\rangle$ to the external particles and is given by 
\begin{equation}
\gamma(n)=t\ \frac{(u\ \alpha',n)}{n!}=\frac{t}{n!}\ \prod\limits_{j=1}^n
[a(u)+j]\sim{(\alpha'u)^n\over n!}\ \ \ ,
\end{equation}
with $a(u)=u\alpha' -1$ the Regge trajectory and $n+1$ the highest possible 
spin of the state $|k;n\rangle$.
As one can see,  the growth of this coupling is not exponential. On the contrary,
although the degeneracy $d_n$ of the Regge states at the level exponentially grows,
\begin{equation}
d_n\sim{\rm const.}~n^{-27/4}\exp(4\pi\sqrt n)\sim\exp(M_n/M_s)\, ,
\end{equation}
 the coupling function $\gamma(n)$ tends to zero for large $n$ due to the $n!$ in the denominator.
Therefore the effective number of Regge states that can be excited  in scattering processes
among the light fields is much smaller compared to $d_n$.




 We now wish to study  how the existence of the string tower affects the evaporation of the 
 semiclassical BHs in string theory.  Of course, for the BHs with the Schwarzschild  radius 
 $r_g \, >  \, l_s$,  the  existence of the string tower  has no effect, since   their production 
 is Boltzmann suppressed.    As already discussed above, such  BHs evaporate into the lowest   excitations and their KK tower.  
 
  The non-trivial question is,  what happens when  BH horizon shrinks to the string size.
Can quasi-classical BHs continue to exist 
 even at sub-stringy distances?  The  exact answer to this question is unknown to us.  However,  as we shall 
 explain, at least from the point of view of an effective theory, there is no obvious inconsistency 
 in continuing semi-classical BH regime beyond the string scale. 
 
    In order to provide an evidence for the above statement,  let us start with a semi-classical BH of size $r_g \, \gg \, l_s$, and follow its evolution.
 We shall assume,  that (by design)  the only non-zero characteristic quantum number of our  BH is 
 its mass ($M_{BH}$).  
In the other words,  by construction,   BH is neutral under all possible massless  gauge fields, with corresponding  charges  that could have  been measured in form of the Gaussian fluxes at infinity.
It also carries no topological or  quantum charge measurable by an Aharonov-Bohm type effect \cite{quantumhair}.     
   Such a BH can always be prepared, for example,  by collapsing a neutral dust cloud with zero angular momentum composed of equal number of particles and anti-particles.  Because 
 the resulting BH is sufficiently large and classical,  by  no-hair  theorem \cite{nohair}, all the possible short distance 
 excitations must vanish outside the horizon.  
     
 Until the BH shrinks to the string length, the evaporation  is semi-classical  and emission takes place 
 into the few lowest massless string states.   At some point,  the BH reaches the size 
 $r_g \, \sim \, l_s$.  Notice,  that in weakly-coupled string theory  (to which we limit our consideration here) 
 the mass of such BH is still much larger than the Planck mass.  This is obvious from the fact that  
 for $r_g \, > \, l_s$ the BH is  semi-classical and  its mass is given by the usual 
 Schwarzschild  relation. For example, in ten dimensions 
 \begin{equation}
 \label{mbh1}
 M_{BH} \, \sim  \, M_{10} (M_{10}\, r_g)^7 \, .  
 \end{equation}
 So,  for  $r_g \, \sim \, l_s$ we get 
  \begin{equation}
 \label{mbh2}
 M_{BH} \, \sim  \, {M_s \over g_s^2} \, \sim \,  {M_{10} \over   g_s^{{7\over 4}}}   \,.  
 \end{equation}
Thus,  for all the accounts, from the point of view of  ten-dimensional Einstein gravity,  such a BH is 
semi-classical.\footnote{For weak string coupling, $g_s<1$, the black hole mass $M_{BH}$ is higher
than the string scale $M_s$ and therefore also higher that the mass of string Regge excitations. E.g.
for $g_s=10^{-2}$ there exist $n_*=g_s^{-2}=10^4$ Regge states before   one reaches the BH mass.}
 Once the BH crosses over  to $r_g \ll l_s$, the semi-classicality can potentially  be 
compromised because of the following two reasons. 

     First, the new heavy fields can emerge outside the horizon and modify the BH metric. 
     Since, by construction, the BH of interest only poses a non-zero $T_{\mu\nu}$ current, 
   only the spin-2 and spin-0 fields may be directly sourced by it.  For example, graviton and dilaton 
   are the simplest examples.   In a ghost-free theory,  gravity mediated by 
  the exchange of spin-2 and spin-0  fields is  always attractive.   Thus,  it is unlikely that such exchanges 
  could lead to either a shrinkage or a  removal of the BH horizon. 
  Of course, the role of the other spins can be enhanced by non-linearities.  However, again, 
  since string theory is weakly coupled,  at least in the closed string theory, the couplings 
  of individual degrees of freedom are suppressed by powers of $M_D$, so it is unclear how the 
  non-linearities could undo the strength of gravity, at least up to the scale, 
  \begin{equation}
  \label{mstar}
  M_* \, = \, {M_D \over N_{eff}^{{1 \over D-2}}} \, ,
 \end{equation}   
   where $M_D$ refers to the $D$ dimensional Planck mass, and  $N_{eff}$ is an effective number 
  of $D$-dimensional degrees of freedom participating in the gravitational  exchanges at scale $M_*$.  The whole question, thus,  boils down  to what is $N_{eff}$?


  In ten dimensions the semi-classical  BH 
 parametrically smaller than  $l_s$ can exist, if  $N_{eff}$ is not exponentially large.  
 This means that the number of effective degrees of freedom affecting 
the BH metric at a given distance,  must  grow slower than the number of the available string states 
with the (naive) Compton wavelength less than that distance.

   The second consideration also indicates that the number of effective degrees of freedom 
  must be less than the actual number of stringy states of a given mass.   The shift of the focus, 
  however, is from the modes affecting the gravitational exchange to the modes participating in the BH evaporation. 
  
  Indeed, when  $r_g \, \ll \,  l_s$
i.e. when the temperature $T \sim  r_g^{-1} \, \gg \,  M_s $, one arrives to an apparent 
conflict with BH semi-classicality:  The number of the energetically-available species in string theory seems to be exponentially large. 

 We shall now show,  that the  key to resolving this puzzle is,  that although the number of species is large, only very few participate in  the evaporation of the BHs. In the other words  $N_{eff}  \, \ll \, N$.

Let us quantify the argument.  As already explained above, it  is a well known fact in string theory  that the density of states grows exponentially fast for $M > M_{s}$. This growth  takes place 
according to (\ref{expnumber}).
The thermodynamics consequences of this exponential growth are well known, and in particular give rise to Hagedorn type transition above the temperatures
$T > M_s$.

Let us now consider an elementary process leading to the BH evaporation in which a semiclassical
nonrotating BH state of temperature $T$ emits, for example,  a closed string state and decays to another nonrotating BH state. If we ignore the backreaction, the BH remains at temperature $T$,
\begin{equation}
\label{act}
BH(T) \,  \rightarrow \, BH(T) \, + \, {\rm String~State} \, .
\end{equation}
This is a good approximation for semiclassical BHs, since, by default,  in semi-classical regime 
the BH is much heavier than its temperature, and the change of temperature due to a single particle emission is a sub-leading effect.    In the leading order, therefore, the temperature can be kept unchanged  for an elementary emission act.   
One might have thought that the exponential growth of the density of states would indicate that the number of available string states for the process~(\ref{act}) is exponentially large, 
however, this is not the case. 

  Let us first show that in string theory there is an absolute upper pound, set by $1/g_s^2$,  on the effective number of Regge states participating in the evaporation of a semiclassical BH. 
 For this, let us first define $N_{eff}$.   Let the one-particle production rate 
 (integrated over all the species) in the evaporation of a semi-classical BH be $\Gamma_{total}(T)$. 
 This quantity is related to the change of temperature as, 
\begin{equation}
\label{tgamma}
{1\over T} {dT \over dt} \, \sim  \,  \Gamma_{total}(T) \, . 
\end{equation}
The violation of semi-classicality simply means that $\Gamma_{total} \, > \, T$. 
For example,  for a ten-dimensional BH evaporating into a single graviton,  we have 
 \begin{equation}
\label{gravitonemission}
\Gamma_{BH \, \rightarrow \, graviton} \sim  T \left({T\over M_{10}} \right)^8 \, , 
\end{equation}
and the BH half-lifetime can be obtained by integrating (\ref{tgamma}).  We can now define 
the $N_{eff}$ as the measure of the relative increase of the evaporation rate with respect  to graviton, 
  \begin{equation}
  \label{defN}
  \Gamma_{total}(T) \, =  \,  \Gamma_{BH \, \rightarrow \, graviton}  \, N_{eff}(T)\, . 
\end{equation}
Obviously,  the effective number of species at the critical temperature,
$\Gamma_{total} (T_c) \, = \, T_c$,  is 
  \begin{equation}
  \label{Ncrit}
  N_{eff}(T_c) \, =  \,  {M_{10}^8 \over T_c^8} \, .  
\end{equation}
Since Regge states only participate in the BH evaporation above the temperature $M_s$, we have 
$T_c \, \gtrsim \, M_s$.  This fact, after taking into the account  the relation between $M_{10}$ and $M_s$,  automatically implies the upper bound on the number  Regge states that can be emitted from a 
semi-classical BHs, 
\begin{equation}
\label{Neffbound}
N_{eff}  \, \lesssim \, {1 \over g_s^2} \, .
\end{equation}

 We shall now study the physics behind this bound.  We shall 
 rely on the effective field theory treatment that explains why the number of states produced in the process~(\ref{act}) is not exponentially large.   
This result also matches  the
general  intuition, that in the decay of a small size object the contribution of the long strings
should be suppressed. It is also in agreement with the general Veneziano-type softening of the stringy amplitudes 
in high-energy scattering processes, discussed earlier. 

For quantifying this result,  we shall use the effective field theory approach of  \cite{scales}, for describing  the closed string emission by external sources, adopting it to the BH evaporation process.
Some results of the analogous study were also briefly reported in \cite{ramy}. 

The key point in our approach is, that from the point of view of an effective theory the process~(\ref{act}) can be described by an effective vertex. The emission of a closed string state
$|a^{\mu_1...\mu_l\mu_{l+1}...\mu_{l+r}}\, p\rangle$ of momentum $p$ by a
black hole of temperature $T$ (and size $r_g=T^{-1}$) has the following form
\begin{equation}
\langle 0| a_{BH}(T')a_{BH}^\dagger(T)\, \alpha_{m_1}^{\mu_1 \dagger}\,...\alpha_{m_l}^{\mu_l \dagger}
\tilde{\alpha}_{\widetilde{m}_1}^{\mu_{l+1} \dagger}\,...\tilde{\alpha}_{\widetilde{m}_r}^{\mu_{l+ r} \dagger} \, |0\rangle,
\end{equation}
where $a_{BH}(T)^{\dagger}$ ($a_{BH}(T)$) is the creation (annihilation) operator for a BH
of temperature $T$,   $\alpha_{m_l}^{\mu_l \dagger}$ ($\alpha_{m_l}^{\mu_l}$) and
$\tilde{\alpha}_{\tilde{m}_r}^{\mu_r \dagger}$ and ($\tilde{\alpha}_{\tilde{m}_r}^{\mu_r \dagger})$ are
the usual string creation (annihilation) operators for the left and right oscillators, respectively.
The $\mu$-s are the Lorentz indexes, and $m_l,m_r$ label the oscillator levels.
Because of the obvious Lorentz structure,
the effective vertex is suppressed by $l\, + \, r\, +\, 2$ powers of $M_{10}$, 
\begin{equation}
 {1\over M_{10}^{l\, +\, r\, +\, 2}} \ p^{\mu_1}\,...p^{\mu_l}
p^{\mu_{l+1}}\,... p^{\mu_{l\, +\, r}}.
\end{equation}
The suppression scale is $M_{10}$ and not $M_s$. 
This is also obvious from the fact that  the closed string states are partners of graviton and couple with gravitational strength.
As a consistency check, this correctly reproduces the graviton vertex, for $l\, =\, r \, =\, 1$, as it should.

Since in BH evaporation, the typical momenta involved are $\sim T$,   the emission rate of a given
string state is suppressed by the factor 
\begin{equation}
\label{rate}
\Gamma_{{\rm single}} \sim  T \left({T\over M_{10}} \right)^{2(l+r)\, + \, 4} \, .
\end{equation}
Before proceeding we shall perform a consistency check, noticing  that the above effective vertex correctly accounts for the decay width of an usual semiclassical BH due to graviton emission. 
Indeed,  the graviton emission rate is (\ref{gravitonemission}). 
This is exactly the rate of the temperature-charge of a semi-classical BH due to evaporation into 
a single graviton, 
\begin{equation}
{1\over T} {dT \over dt} \, \sim  \,  {T^9 \over M_{10}^8} \, . 
\end{equation}
The half life-time of such a BH can be estimated by integrating the above equation, which gives 
\begin{equation}
\label{lifetime} 
\tau_{BH} \, \sim \,  {M_{10}^8  \over T^9} \, .
\end{equation}
This matches the lifetime obtained by the usual consideration 
of a BH as of a perfect quantum emitter.  In the absence of any species other than graviton, the change of BH mass would be given by,    
\begin{equation}
{d M_{BH} \over dt} \, \sim  -  \, r_g^{8} T^{10} \, \sim \, - T^2 \, .
\end{equation}
Integrating this equation, we get exactly (\ref{lifetime}). 

  In order to obtain the total one particle emission rate we have to sum (\ref{rate}) over all the 
  states with masses up to $T$, or equivalently up to the level $n \, = \, T^2/M_s^2$. 
  Although the number of states is exponentially large,  what matters is the number of states with 
  {\it fixed} oscillator number $l+r$.  Finding this number reduces to a combinatorics problem
  of partitions $p_{l+r}(n)$ of an integer $n$, which for  $l+r \, \ll \, n$ scales as 
  \begin{equation}
  p_{l+r}(N) \, \sim \, {(n\, - \, (l+r))^{l+r -1} \over (l+r)! (l+r-1)!} \, .
  \end{equation}  
Thus, the total number of Regge states up to the level $n$ that are created by $l$ left and $r$ right oscillators grows with $n$ as 
 \begin{equation}
  p_{l+r}^{(d)}(N) \, \sim \, n^{l+r -2} \, d^{l+r} \, .
  \end{equation}  
Thus, the total one particle emission rate into closed string states from a BH of temperature 
$T$ is, 
\begin{equation}
\label{ratetotal}
\Gamma_{{\rm total}} \sim  \sum_{r+l} \, \sum_{n}^{n_{max} = T^2/M_s^2} \, T \left({T\over M_{10}} \right)^{2(l+r)\, + \, 4} \, n^{l+r -2} \, d^{l+r} \, .
\end{equation}
Performing summation over $n$ we get
\begin{equation}
\label{ratetotal1}
\Gamma_{{\rm total}} \sim  \, T \left({T\over M_{10}} \right)^{6} 
 \sum_{r+l} \,\left(\sqrt{d} {T^2\over M_{10} M_s} \right) ^{2(l+r) -2} \, 
\end{equation}
This expression clearly indicates the physical reason behind the suppression of the effective number of species.  
 For example,  till the temperature  $T_{(2)} \, \equiv \sqrt{M_{10}M_s} \, = M_s/g_s^{1/8}$.
the rate is dominated by $l+r=2$.
Therefore, we have
\begin{equation}
\Gamma_{total}(T <  T_{(2)}) \, \sim \,  \Gamma_{(2)} \, \equiv \ T \left({T\over M_{10}} \right)^{8}\, \left({T\over M_s} \right)^{2} \,.
\label{2-rate}
\end{equation}
Thus, up to the temperature $T \,  \sim \, \, M_s/g_s^{1/8}$, the only states that participate in the process~(\ref{act}) are the two-oscillator states. These include the spin-$0$ and spin-$2$ states only. Thus, the effective number of species contributing to the evaporation of a neutral nonrotating BHs at temperatures $T \ll T_{(2)} $ is 
\begin{equation}
N_{eff} \sim \left({T_{(2)} \over M_s} \right)^{2} \, \sim \, g_s^{-1/4}\, ,
\end{equation}
despite the fact that the total number of thermally-available species is exponentially large. 

Although, the (\ref{2-rate}) breaks down above $T_{(2)}$, above which series have to be resumed, it serves the purpose illustrating that the number of  string species participating in the BH evaporation 
at temperatures $\gg \, M_s$  is {\it not} exponentially large, in accordance with (\ref{Neffbound}).  
 In the next section we shall provide another example of BH evaporating in the string states, in which relation  (\ref{Neffbound}) is explicit.

 When extending the above analysis to the open string states, the following subtlety appears. 
  
   First,  the perturbative open string states  live on a lower dimensional sub-manifolds, so for them 
   we have to take into the account the non-democracy factor in BH evaporation. 
 
  Secondly,  the open string states may inter-couple without 
 any $M_{10}$-suppressed interactions.  Such unsuppressed couplings to the  semi-classical BHs
 are impossible, because the evaporation strength has to be controlled by gravity.     
  Consider, for example,  a single photon emission from a neutral BH.   The fact that emission must be controlled by the gravitational strength implies,  that the effective vertex through which the emission of photon from a neutral BH takes place must be of a dipole type,
  \begin{equation}
  \label{photon}
      F_{\mu\nu}  \,  J^{\mu\nu}_{BH}  \, , 
  \end{equation} 
  where $J^{\mu\nu}_{BH}$ is the BH thermal current, which classically should vanish for any neutral non-rotating BH, by no-hair theorem \cite{nohair}.    This form of the vertex is also dictated by the gauge invariance, since  the BH is electrically neutral.
  
    The $M_{10}$ dependence of $J_{BH}^{\mu\nu}$ must be such that to reproduce the correct  species number dependence of the lifetime of a semi-classical BH that evaporates into photons and gravitons, 
 \begin{equation}
\label{photonemission}
\Gamma_{BH \, \rightarrow \, \gamma} \sim  T \left({T\over M_{10}} \right)^8 \, . 
\end{equation} 
The above is of course very different  from the  photon emission rate from an ordinary relativistic hot plasma composed of, say, 
open string zero modes (e.g., electron positron), the photon emission rate out of which can be estimated as  $\sim \, \alpha_{EM}  T$. 

\section{Physical Analogy: Black Hole Evaporation in QCD Strings} 

 It is a well known idea \cite{largeN} that at large distances $SU(n)$ QCD without light quarks 
 can be described as a theory of closed strings. The strings in question are cromo-electric  
flux tubes, with the tension $\sim \, \Lambda_{QCD}$.  In the presence of quarks, the QCD closed 
strings can break into the open ones, that connect quarks-antiquark pairs.  

 The suggested connection between the QCD and string theory becomes more apparent  in large 
 $n$ limit. In this limit QCD flux tubes interact as fundamental  strings with the effective  string coupling $g_s \, \equiv \, {1 \over n}$ and the corresponding string scale $M_s \, \equiv \, \Lambda_{QCD}$. 
         At distances  $ \ll \Lambda^{-1}_{QCD} \, \equiv \, l_{QCD}$ such a string theory becomes a gauge theory  of  $N \, \simeq \, n^2 $ gluon species.  
 
   We shall now consider the $SU(n)$-QCD coupled to Einsteinian gravity, and investigate evaporation of 
   semi-classical black holes in such a theory.  Due to obvious reasons, we shall limit our analysis to four
   dimensions.
    Since we are primarily interested in the evaporation of the  small quasi-classical black holes with the size $R \, < \, l_{QCD}$,  we shall  assume the hierarchy 
  ${M_4 \over n}  \, \gg \, \Lambda_{QCD}$. 
  
    Although in no way we can claim a full physical equivalence between the above system and the 
  semiclassical BH evaporation in a ``real'' string theory,  nevertheless,  it enables to grasp in more controllable way the  important general properties of the system in which BHs can evaporate into the stringy objects.  The advantage of the QCD string example is,  that it admits the short-distance 
description in terms of  $n^2$ gluon  degrees of freedom, and thus,  allows for the cross-check of the results 
  obtained within the  effective string-black hole interaction picture.    As we shall see, the above  complementary (gluon) description qualitatively 
 confirms our results obtained in the previous section.  Although naively the number of thermally-available string resonances (in QCD-string case, glueballs) is exponentially large, the effective number of species to which the black hole evaporation is taking place is much less, and is given by the number 
 of gluons, rather than glueballs.  
    
  Let us consider the evaporation of a BH in such a theory.  The  large BHs, with size $\gg 
  l_{QCD}$,  cannot evaporate into the QCD degrees of freedom and are uninteresting for our purposes. 
  We shall thus focus on the black holes  with size $l_*  \, \ll \, R \, \ll  l_{QCD}$.  Such black holes are 
  semi-classical and undergo the usual Hawking evaporation. They are much hotter than the  QCD scale, and thus,  must be able to evaporate into the QCD strings states (glueballs). 
  The existence of the two descriptions gives us an useful tool for understanding  the evaporation 
 rate. 
 
   For this,  consider the evaporation process described by the two observers.  One is a microscopic 
   observer  operating at distances $\ll  \, l_{QCD}$ and describing the black hole evaporation in terms of 
   the gluon emission.   Another is a long-distance observer that is using the effective description at distances  $\gg l_{QCD}$.  The latter observer cannot resolve the 
 structure of the QCD strings,  and can only  describe the black hole evaporation process in terms of 
 closed string (glueball) emission.   
 
   The black hole evaporation rates obtained by the two observers must agree, since the two descriptions are complementary. 
   Of course, the gluons observed by the short-distance observer will undergo the process of 
   glueballization  after the QCD time  $t_{QCD} \sim l_{QCD}$, but the details of this process are unimportant for the 
   large-distance description. The macroscopic  observer interested in detecting evaporation process 
   at large distances, must be able to integrate out the microscopic physics and understand the 
   glueballization process in terms of the effective interaction vertex between the black hole 
   and the closed string states. 
     This is the key point of the effective long-distance description.    
     
      Thus, if large-$n$ QCD can indeed be approximated as a closed string theory, the general structure of the black hole - glueball interaction must be understandable  in form of the effective black hole -string 
      interactions similar to the one considered in the previous section.

  The complementarity of the two descriptions then allows for the effective number of closed strings (glueballs) participating in the emission process to be understandable  in  terms of the number of gluons. 
 The microscopic theory of black hole - gluon evaporation, then immediately  puts the limit on the 
number of  macroscopic closed string states produced.  In microscopic description the black hole 
 evaporates in $N^2$ elementary gluon states, and hence its evaporation rate satisfies the 
 quasi-classicality condition till the temperature $\sim M_*$, despite the fact that the naive 
 number of thermally-accessible  glueball states may be much larger.           
  
  Translating in terms of  the string coupling, we obtain that the effective number of species participating in black hole evaporation in QCD string theory is $N_{eff} \, \sim \, 1/g_s^2$, as opposed to being 
  set by the number of glueballs!  
  It is rather tempting to generalize this number to the ordinary strings.  Of course, in the 
 usual string theory no QCD type complementary description is known that would enable us to think of fundamental strings as of flux-tubes of an underlying theory with finite number of elementary degrees of freedom.  However,  for our analogy, the absence of such description is unimportant, since the effective black hole -string interactions must be qualitatively insensitive to such descriptions, and should only rely on the stringy nature of the macroscopic description.  
    
    For  QCD strings, existence of the complementary short-distance description gives simple microscopic explanation  to the fact that the long closed strings are produced very selectively. The long strings that are produced more efficiently are the 
    ones that are the results of the glueballization process of a single gluon emission from the black hole.  
    At temperature $T$ the maximal length of such flux tubes is $l_{max} \, \sim {Tl_{QCD}^2}$, but not all the possible tube configurations of this length are produced.   Thinking in terms of the different harmonics of the macroscopic oscillating flux-tubes, the strings that can be produced efficiently are the 
 ones that during their oscillation undergo contraction to the BH size of $T^{-1}$.

\section{Conclusions}

 In this paper we have studied the question of how the string theory compactifications accommodate the BH bound (\ref{bhbound}).
 The bound imposes a strict condition on the gravitational cutoff of the theory
in terms of number of species to which the semi-classical BHs can evaporate in any consistent theory of gravity.
  The issue is  therefore  dramatically linked to the questions, down to what critical size the semi-classical BHs can exist in string theory, and to what number of particle species they can efficiently evaporate?

  We have divided our studies into the cases of  BHs larger or smaller than $l_s$. 
 The first category of the BHs can only evaporate into the open and closed string zero modes and their KK resonances. We have shown that the number
of species to which such BHs can evaporate automatically saturates the
bound (\ref{bhbound}).  In fact, saturation of this bound becomes equivalent
to the geometric relations obtained between the four-dimensional Planck mass and
the string scale in terms of the geometry of the internal manifold.

 We thus discover, that the bound allows for understanding of seemingly-complicated fundamental geometric relations in terms of simple counting of the number of species.

 We next  considered the question of sub-stringy BHs.
 The naive intuition tells us, that such BHs should not exist based on
thermal arguments a la Hagedorn. Since the number of string Regge states
increases exponentially above $M_s$, the semi-classicality condition
(\ref{Tchange}) should be violated almost instantly as soon as the BH
size crosses $l_s$. We showed that this intuition is false, and the thermal arguments do not prevent the semi-classical BHs from shrinking down to
sub-stringy size.  The key reason is, that despite the exponentially growing number of string oscillator states, only  few are produced in the evaporation of the small BHs.  

  Our study suggest that the effective number of string species is
given by $N_{eff} \, = \, 1/g_s^2$. Interestingly, the BH  bound (\ref{bhboundD}) applied to this number exactly reproduces the relation between the string
and Planck masses. This remarkable fact suggest an yet to be understood fundamental connection between the particle species and gravity. 

  Our findings have obvious phenomenological implications for TeV scale string
scenario, motivated by the solution of the hierarchy problem.
 We predict that the maximal effective number of species to which the micro BHs can decay must be given by $1/g_s^2$.

\section*{Acknowledgements}
We like to thank C. Gomez,  F. Quevedo and S. Stieberger for useful discussions.  
The work of G.D is supported in part
by European Commission  under 
the ERC advanced grant 226371,  by  David and Lucile  Packard Foundation Fellowship for  Science and Engineering and  by the NSF grant PHY-0758032. 
 The work of D.L. is supported by the Excellence Cluster
{\sl Origin and Structure of the Universe} in Munich.

\end{document}